\documentclass[a4paper,12pt,titlepage]{article}

\usepackage{amsmath,amssymb,revsymb}
\usepackage{here}
\usepackage{bm}
\usepackage{fancybox}
\usepackage{secdot}
\usepackage{comment}
\usepackage{braket}
\usepackage{titlesec}
\usepackage{geometry}
\usepackage{hyperref}
\usepackage{bookmark}
\newcommand{\slashed}[1]{{\ooalign{\hfil/\hfil\crcr$#1$}}}
\usepackage{fancyhdr}
\usepackage{boxedminipage}
\usepackage{color}
\usepackage{feynmf}
\usepackage{multirow}

\makeatletter
\renewcommand{\theequation}{%
\thesection.\arabic{equation}}
\@addtoreset{equation}{section}
\makeatother

\makeatletter
\renewcommand{\thetable}{%
\thesection.\arabic{table}}
\@addtoreset{table}{section}
\makeatother

\titleformat*{\section}{\LARGE\bfseries}
\titleformat*{\subsection}{\Large\bfseries}

\hypersetup{
setpagesize=false,
bookmarksnumbered=true,%
bookmarksopen=true,%
colorlinks=true,%
linkcolor=blue,
citecolor=blue,
}

\begin{document}


\fancypagestyle{foot}
{
\fancyfoot[L]{$^{*}$E-mail address : hinata0903@akane.waseda.jp }
\fancyfoot[C]{}
\fancyfoot[R]{}
\renewcommand{\headrulewidth}{0pt}
\renewcommand{\footrulewidth}{0.5pt}
}

\renewcommand{\footnoterule}{%
\kern -3pt
\hrule width \columnwidth
\kern 2.6pt}


\begin{titlepage}
\begin{flushright}
\begin{minipage}{0.2\linewidth}
\normalsize


WU-HEP-20-07\\*[50pt]
\end{minipage}
\end{flushright}

\begin{center}

\vspace*{5truemm}
\Large
\bigskip\bigskip

\LARGE\textbf{Mass hierarchy from the flavor symmetry in supersymmetric multi-Higgs doublet model}%

\Large

\bigskip\bigskip
Atsushi Hinata$^{*}$%
\vspace{1cm}

{\large \it{Department of Physics, Waseda University, Tokyo 169-8555, Japan}}\\
\bigskip\bigskip\bigskip

\large\textbf{Abstract}\\
\end{center}
\ \ We study the supersymmetric standard model with multiple Higgs doublets with gauged $\mathrm{U}(1)_X$ flavor symmetry. When the flavor symmetry is broken by the vacuum expectation value of flavon, the $\mathbb{Z}_3$ symmetry $M_3$ called {\it matter triality} remains and it prohibits the baryon number violation up to dimension-5 operators. We study the contribution of the extra-Higgs fields to the anomaly cancellation of flavor symmetry and analyze the mass spectra including the multiple generations of Higgs fields as well as quarks and leptons. We show a series of $\mathrm{U}(1)_X$ charge assignments, which reproduce the observed masses and mixing angles of quark and lepton. We also find that, with such realistic charge assignments, the extra-Higgs fields obtain masses around the intermediate scale and decouple from the electroweak physics because of the holomorphy of superpotential.

\thispagestyle{foot}

\end{titlepage}

\baselineskip 7.5mm


\tableofcontents
\clearpage


\parindent=30pt

\section{Introduction}
\ \ The standard model (SM) in particle physics has successfully explained the current experimental results including the discovery of the Higgs boson at the LHC. However, there are unsolved problems, such as the fine-tuning problem to explain the Higgs boson mass, the absence of the dark matter candidate and so on. Supersymmetry is one of the most plausible candidates which can stabilize the hierarchy between the Planck scale $M_{Pl} \sim 2.4 \times 10^{18}{\mathrm{GeV}}$ and the electroweak (EW) scale.
The simplest supersymmetric extension of the SM, the minimal supersymmetric standard model (MSSM) provides a viable phenomenology by assigning an Abelian discrete symmetry, called $R$-parity \cite{Martin:1997ns}. Due to this symmetry, the baryon and lepton number violations are prohibited at the renormalizable level, thus the lightest supersymmetric particle can be the most attractive candidate for the dark matter and the proton decay mediated by the superpartner is also suppressed. In spite of those favorable features, however, the baryon or lepton number violations arise from higher-dimensional operators, those can cause problems even if they are suppressed by $M_{Pl}$ \cite{Goto:1998qg, Murayama:1994tc}. In this context, alternative discrete symmetries embedded into some continuous gauge symmetries have been investigated by some previous works \cite{Ibanez:1991pr,Ibanez:1991hv, Dreiner:2005rd, Lee:2007qx, Higaki:2019ojq, Abbas:2017vws, Abbas:2018lga}, based on the argument that the global symmetries are broken by the gravitational effects \cite{Krauss:1988zc, Kallosh:1995hi}. It is known that not only $R$-parity or baryon triality but also matter triality $M_3$ \cite{Lee:2007qx} is consistent with the gauge theory from the viewpoint of discrete anomaly cancellation \cite{Ibanez:1991pr, Araki:2008ek} when we consider the three generations of right-handed neutrino.

On the one hand, the fermion mass hierarchy has been also studied as one of the open problems in SM. Although three generations of SM fermions have the same quantum number, those masses are different from each other and the gap between the generations is extremely large. In addition to this hierarchy, the mixing pattern of lepton is different from quark mixing. One of the promising explanations of such hierarchies is the flavor symmetry. Many flavor models have been proposed to realize the mass hierarchy or the mixing of quark and lepton \cite{Davidson:1979wr, Davidson:1979db, Davidson:1979nt, Ishimori:2010au, King_2013}. 
In particular, the Froggatt-Nielsen (FN) mechanism \cite{Froggatt:1978nt} can realize the observed mass spectra and flavor mixings. In this mechanism, the higher-dimensional operators generate hierarchical Yukawa matrices, where an additional scalar field called flavon is introduced, which is charged under $\mathrm{U}(1)$ symmetry. The flavor symmetry distinguishing the generation is spontaneously broken by a non-vanishing vacuum expectation value of the flavon. In \cite{Dreiner:2003hw, Harnik:2004yp, Dreiner:2006xw, Dreiner:2007vp}, by gauging the flavor symmetry denoted here $\mathrm{U}(1)_X$, it is argued that certain discrete symmetries prohibiting the proton decay can be obtained as a subgroup of $\mathrm{U}(1)_X$ after its spontaneous breaking. However, the realization of FN-mechanism with a proper discrete symmetry requires a highly fractional charge assignment to cancel out the anomalies. Naively, the strict constraints from the anomaly and phenomenological requirements such as fermion mass hierarchies will be relaxed by the addition of a new flavor-charged field. Therefore, we suggest the existence of multiple generations of the Higgs field as a reasonable extension. The extended Higgs sector is also favored from the viewpoint of UV-theory, because the extra-Higgs fields are sometimes inevitable in the context of string compactification, for example, orbifold model \cite{Font:1989aj}, or magnetized orbifold model \cite{Aldazabal:2000sa}. On the other hand, the phenomenological aspect of multi-Higgs doublet models (MHDMs) has been discussed \cite{Branco:2011iw, Darvishi:2019dbh, Ivanov:2007de, Aranda:2000zf, Georgi:1978ri, Shanker:1981mj, McWilliams:1980kj, Escudero:2005hk} regardless of the existence of SUSY. From the viewpoints of the low energy physics, the additional scalar states induce the flavor changing neutral currents (FCNCs) which affect $K$, $B$, and $D$ meson mixing at tree level, thus those masses are strictly constrained \cite{Escudero:2005hk}.

In this paper, we study the supersymmetric SM in MHDM with the $\mathrm{U}(1)_X$ flavor symmetry. The spontaneous breaking of gauged $\mathrm{U}(1)_X$ symmetry respects the discrete symmetry $M_3$, thus the baryon number violating operator is prohibited up to dimension-5 operators. The flavor symmetry respecting $M_3$ can be anomaly free by introducing three generations of right-handed neutrino and we confirm that the relevant anomalies are canceled out by the Green-Schwarz mechanism \cite{Green:1984sg}. Yukawa hierarchy can be realized via the Froggatt-Nielsen mechanism by suitably choosing the flavor charge. The charge assignment also determines the structure of $\mu$-matrix which is a mass matrix for multiple Higgs fields in the superpotential. By choosing the specific charge, the extra-Higgs field can be decoupled and the MSSM-like Higgs field which is responsible for the EW symmetry breaking remains at the low energy. We perform a numerical search and show that there are such flavor charges, which effectively realize MSSM without extra-Higgs fields below the intermediate scale.

The organization of this paper is the following. In Section \ref{sec:model}, we show the model set-up about the discrete symmetry $M_3$ and the anomaly cancellation condition. Then, in Section \ref{sec:hie}, we confirm the matter sector including fermion and Higgs fields. From the experimental points of view, observed masses and mixing angles of quark/lepton constrain the flavor charge. We introduce the non-minimal coupling in K\"ahler potential accommodating Giudice-Masiero mechanism \cite{Giudice:1988yz}, and show that the decoupling of extra-Higgs fields can be realized. In Section \ref{sec:res}, we show the concrete examples of charge assignment and the numerical evaluation of the relevant observables. Finally, we conclude in Section \ref{sec:con}.

\section{The SUSY SM with matter triality}
\label{sec:model}
\ \ Firstly, we will confirm our model set-up. In the notation of supersymmetric theory, the chiral superfields are $\Phi_i=Q_i, \bar{U}_i, \bar{D}_i, L_i, \bar{E}_i, \bar{N}_i, H_u, H_d$ and those lower indices for matter fields indicate each generation ($i=1\sim 3$). The right-handed neutrinos are introduced as the extension of the fields contents.

The anomaly-free discrete symmetries have been discussed in the context of proton stability. Proton lifetime is strictly bounded \cite{Tanabashi:2018oca} and its partial mean life have been measured $\tau_p > 1.6 \times 10^{34} \ {\mathrm {years}}$ for lepton channel $p^+\rightarrow e^+\pi$. Due to the existence of the supersymmetric partner of the SM fields, additional proton decay processes by the baryon or lepton number violating operators are also predicted in supersymmetric SM (SSM). Furthermore, the higher dimensional operators, which cannot be prohibited by $R$-parity, contribute to the dangerous process. We list the baryon or lepton number violating operators up to dimension-5 in Appendix \ref{sec:BLVop}.

Ib\'a\~nez and Ross proposed the Abelian discrete symmetry instead of $R$-parity \cite{Ibanez:1991pr, Ibanez:1991hv}. According to \cite{Krauss:1988zc, Kallosh:1995hi}, the quantum gravity effect violates the global symmetry regardless of continuous or discrete one. In this context, to be consistent with the argument, they consider the global discrete symmetry should be embedded into gauge symmetry, and it is the remnant of spontaneous breakdown, {\it i.e.}, $\mathrm{U}(1)_X \rightarrow \mathbb{Z}_{2,3}$.
In this paper, we adopt a particular ${\mathbb{Z}}_3$ symmetry proposed in \cite{Ibanez:1991pr, Lee:2007qx}, it is called {\it matter triality}. It requires the three generations of the right-handed neutrino in order to cancel the gauge anomaly. Under this symmetry, each matter field transforms as Table \ref{tb:discrete symmetry}, where $\omega=e^{2\pi i /3}$.

\begin{table}[htb]
\centering
\caption{The representation of charged particle under matter triality.}
\begin{tabular}{c|cccccccccc}\hline\hline
& $Q$ & $\bar{U}$& $\bar{D}$& $L$& $\bar{E}$& $\bar{N}$ &$H_u$ &$H_d$\rule[-0pt]{0pt}{12pt} \\ \hline
$M_3$ & $1$ & $\omega^2$& $\omega$& $\omega$& $1$& $\omega$ &$\omega$ &$\omega^2$ \rule[-0pt]{0pt}{12pt}\\ \hline\hline
\end{tabular}
\label{tb:discrete symmetry}
\end{table}

The superpotential under matter triality is given by
\begin{align}
W
=y_{ij}^u\bar{U}_iQ_jH_{u} +y_{ij}^d\bar{D}_iQ_jH_{d}+y_{ij}^e\bar{E}_iL_jH_{d} +y_{ij}^\nu\bar{N}_iL_jH_{u} + \mu H_{u}H_{d} + \lambda_{ijk}^\nu\bar{N}_i\bar{N}_j\bar{N}_k.
\label{eq:effectivesuperpotential}
\end{align}
Note that the introduction of the right-handed neutrino leads to the interaction term in Eq.(\ref{eq:effectivesuperpotential}) which violates the lepton number. However, the baryon number is conserved up to mass dimension-5, thus, the proton stability is ensured by matter triality.

\subsection{Multi-Higgs doublet model}
While the Higgs field has been discovered at the LHC, the multi-Higgs doublet models (MHDMs) are predicted from some UV-theories. For example, since the down-type Higgs field belongs to the gauge groups with the lepton doublet, then the multiple Higgs fields appear in the string compactification \cite{Font:1989aj, Aldazabal:2000sa, Otsuka:2018rki}. Thus, it is meaningful to consider the extended Higgs sector. On the other hand, the decoupling of the extra-Higgs fields is also one of the main issues in MHDMs as we mentioned above.

Let us introduce the multiple Higgs fields $H_u\rightarrow H_{u\alpha}$ and $H_
d\rightarrow H_{d\alpha}$, which belongs to the same SM gauge groups. The Greek index of the Higgs fields run from $1$ to $N_H$. Those fields contribute to the anomaly cancellation condition (the detailed discussion of the gauge anomaly cancellation is given in the following Section \ref{sec:ano}). By introduction of the extra fields, the interaction containing the Higgs fields are extended ($y_{ij}^u\rightarrow y_{ij\alpha}^u$, $\mu\rightarrow \mu_{\alpha\beta}$). Note that there remains an ambiguity to choose the discrete charge for the extra-Higgs fields. If those have the same discrete charge with the first generation, the potential minimization of the EW vacuum is necessary to analyze the full Higgs potential due to the Yukawa coupling for the extra fields. On the other hand, if not, the additional baryon/lepton number operators with the Higgs fields appear in the superpotential.

\subsection{Anomaly cancellation condition}
\label{sec:ano}
The gauge anomaly cancellation requires that the anomaly coefficients ${\cal{A}}_{\cdots}$ must be canceled. Those coefficients ${\cal A}_{\cdots}$ are evaluated as the algebraic equation of $X$-charges according to Fujikawa methods \cite{Fujikawa:1979ay}. We must consider the SM gauge anomaly $G_{SM}=\mathrm{SU}(3)_C\times \mathrm{SU}(2)_L\times\mathrm{U}(1)_Y$, and pure $\mathrm{U}(1)_X$ anomaly and those coefficients are evaluated by the flavor charge:
\begin{eqnarray}
{\cal A}_{CCX} &=& \sum_i[{2X_{Qi} +X_{Ui}+X_{Di}}],\label{eq:CCX}\\
{\cal A}_{WWX} &=& \sum_i[{3X_{Qi} +X_{Li}}]+\sum_\alpha(X_{H_u\alpha}+X_{H_d\alpha}),\label{eq:WWX}\\
{\cal A}_{YYX} &=& \frac{1}{6}\sum_i[X_{Qi}+8X_{Ui}+2X_{Di}+3X_{Li}+6X_{Ei}] \nonumber \\
&& \ \ \ \ +\frac{1}{2}\sum_\alpha(X_{Hu\alpha}+X_{Hd\alpha}),\label{eq:YYX} \\
{\cal A}_{YXX} &=& \sum_i[X_{Qi}^2-2X_{Ui}^2+X_{Di}^2-X_{Li}^2+X_{Ei}^2] \nonumber \\
&& \ \ \ \ +\sum_\alpha(X_{Hu\alpha}^2-X_{Hd\alpha}^2), \label{eq:YXX} \\
{\cal A}_{XXX} &=& \sum_i[X_{Qi}^3+X_{Ui}^3+X_{Di}^3+X_{Li}^3+X_{Ei}^3] \nonumber \\
&& \ \ \ \ +\sum_\alpha(X_{Hu\alpha}^3+X_{Hd\alpha}^3) + {\cal A}_{XXX}^{exotic} \label{eq:XXX} ,
\end{eqnarray}
where ${\cal A}^{exotic}_{XXX}$ is the contribution from exotic fields assigned $\mathrm{U}(1)_X$ charge in hidden sector. Note that the consistency of gauge theory requires that all anomaly coefficients must be canceled. Even if the anomaly-free discrete symmetry is assigned, it is necessary to choose the charge assignment of $\mathrm{U}(1)_X$. We require the anomaly cancellation by Green-Schwarz (GS) mechanism \cite{Green:1984sg, Binetruy:SUSY}. Let us assume that the single GS field charged under $\mathrm{U}(1)_X$. Then, the string axion appears in the gauge sector as
\begin{equation}
{\cal L} = -\frac{1}{4}s(x)\sum_i k_i F_{i\mu\nu}F_i^{\mu\nu} +\frac{1}{4}a(x)\sum_i k_i F_{i\mu\nu}\tilde{F}_i^{\mu\nu},
\end{equation}
where $s(x)$ and $a(x)$ are the dilaton and axion fields. The normalization factor $k_i$ is the affine/Ka\v{c}--Moody level, and the field strength $F_{i\mu\nu}$ (and its dual $\tilde{F}_{i\mu\nu}$) are given for corresponding gauge groups, thus $i$ runs over $G_{SM}\times\mathrm{U}(1)_X$. Under the $\mathrm{U}(1)_X$ gauge transformation ($A_\mu^X\rightarrow A_\mu^X -\partial_\mu\theta(x)$), the anomaly coefficients appears as,
\begin{equation}
\delta{\cal L} =-\frac{1}{8}\theta(x)\sum_i {\cal{A}}_iF_{i\mu\nu}\tilde{F}_i^{\mu\nu} ,
\end{equation}
where the anomaly coefficients ${\cal A}_i$ correspond to the Eq.(\ref{eq:CCX}), (\ref{eq:WWX}), (\ref{eq:YYX}), and (\ref{eq:XXX}). The dilaton field which has shift symmetry under $\mathrm{U}(1)_X$ symmetry compensates the deviation $S\rightarrow S + i\delta_{GS}\Lambda_X/2$, where $\Lambda_X$ is gauge transformation parameter and $\delta_{GS}$ is determined by the anomaly coefficients. Therefore, the Lagrangian is invariant if the anomaly coefficients satisfy the following relation,
\begin{equation}
\frac{{\cal{A}}_{CCX}}{k_C}=\frac{{\cal{A}}_{WWX}}{k_W}=\frac{{\cal{A}}_{YYX}}{k_Y}=\frac{{\cal{A}}_{XXX}}{k_X}=\delta_{GS}.
\end{equation}
While the normalization of the non-Abelian symmetry is restricted in integer, the normalization of the Abelian symmetry cannot be determined by algebraic way. If the SM gauge groups are unified into the simple group, the Ka\v{c}-Moody levels of the SM gauge group can be related by the standard GUTs normalization \cite{Ginsparg:1987ee},
\begin{equation}
k_C=k_W=3k_Y/5.
\end{equation}
On the other hands, in the heterotic string theory, the hypercharge normalization can be determined by the decomposition of the gauge group \cite{Aldazabal:2000sa, Otsuka:2018rki}. In particular, in $\mathrm{SO}(32)$ heterotic string theory, the one-loop threshold correction to the gauge coupling for non-Abelian gauge group are non-universal \cite{Blumenhagen:2005pm, Abe:2015xua}, thus the gauge coupling unification depends on the correction even if the normalization is not canonical one. In our calculation, let us assume that those of the non-Abelian gauge group is one and the hypercharge normalization $k_Y$ is the parameter to solve the anomaly cancellation condition\footnote{The gauge coupling unification in the context of non-standard hypercharge normalization is discussed in Refs.\cite{Dienes:1996du, PerezLorenzana:1999tf, Schwichtenberg:2018cka}.}.
Therefore, the X-charge and $k_Y$ must satisfy the following relation,
\begin{equation}
{\cal A}_{CCX}={\cal A}_{WWX}, \ \ {\cal A}_{WWX}={\cal A}_{YYX}/k_Y.
\end{equation}
Another hypercharge anomaly ${\cal A}_{YXX}$ cannot be absorbed by the shift of dilaton, therefore we require additional condition, ${\cal A}_{YXX}=0$. Note that the pure $\mathrm{U}(1)_X$ anomalies are affected by the exotic sector, thus it is necessary to assume the concrete form of exotic sector. Thus, we omit this constraint in the following discussion\footnote{The gravitational anomaly has to be also considered, but we omit it in our analysis since the hidden sector contributes to the anomaly cancellation condition.}.

\section{Mass hierarchy of fermion and Higgs sector}
\label{sec:hie}
\subsection{Fermion mass hierarchy}
\ \ Fermion mass hierarchy is one of the open problems in SM. Quark and lepton have the generation structure, which have the hierarchical mass spectra, and those masses are evaluated by the Cabbibo angle $\epsilon \sim 0.22$,
\begin{eqnarray}
m_u : m_c : m_t &\sim& \epsilon^8 : \epsilon^4 : 1, \nonumber \\
m_d : m_s : m_b &\sim& \epsilon^4 : \epsilon^2 : 1, \\
m_e : m_\mu : m_\nu &\sim& \epsilon^{4,5} : \epsilon^2 : 1 \nonumber .
\end{eqnarray}
To explain the large hierarchy between the masses of generations, we will adopt the Froggatt-Nielsen mechanism \cite{Froggatt:1978nt} and identify the gauge symmetry $\mathrm{U}(1)_X$ with flavor symmetry. This flavor symmetry respects the matter triality, therefore each charged field transforms as same discrete charge with respect to the generation after $\mathrm{U}(1)_X$ symmetry breaking. Above the flavor symmetry breaking scale, the superpotential Eq.(\ref{eq:effectivesuperpotential}) is modified to be gauge invariance. For example, the Yukawa coupling for up-type quark can be obtained from the higher dimensional operator,
\begin{equation}
g_{ij\alpha}^u\Theta[n_{ij\alpha}^u]\left(\frac{A}{M_{Pl}}\right)^{n_{ij\alpha}^u}\bar{U}_iQ_jH_{u\alpha} ,
\label{eq:WYU}
\end{equation}
where $A$ is the flavon superfield and it has the X-charge $X_A=-3$. The coupling constants $g_{ij\alpha}^u$ are assumed as ${\cal O}(1)$ {\it i.e.}, $\sqrt{\epsilon}\leq g_{ij\alpha}^u \leq 1/\sqrt{\epsilon}$.
$\Theta[x]$ is equal to $1$ for $x\leq 0$ or $0$ for others. The other operators can be rewritten under this flavor symmetry. After the flavor symmetry breaking, the effective operators Eq.(\ref{eq:effectivesuperpotential}) are obtained. Then, the vacuum expectation value of flavon generates the hierarchical structure by the ratio $\braket{A}/M_{Pl}=\epsilon$.
In order to obtain the correct mass spectrum of quark and lepton, let us consider the specific ansatz for the Yukawa matrices.
Note that the extra-Higgs fields can be mixed in the diagonalization of the mass matrix and its effect is possible to contribute to the Yukawa matrix, but for simplicity, let us assume that the first generation of the Higgs fields only contributes to the Yukawa coupling. This assumption is justified by the requirement of decoupling of the extra-Higgs fields. For the Yukawa coupling of up-type quark {\it i.e.}, the ansatzes of the $\epsilon$-suppression is given by
\begin{equation}
n_{ij1}^u =
\begin{pmatrix}
8 & 7-y & 5-y \\
5+y & 4 & 2 \\
3+y & 2 & 0
\end{pmatrix}_{ij},
\label{eq:YUansatz}
\end{equation}
where $y=0,1$ is an integer parameter related to the CKM mixing matrix.
For the down-type quark and charged electron, those ansatzs can be also written in terms of some parameters $x, y, z$,
\begin{equation}
n_{ij1}^d =
\begin{pmatrix}
4+x & 3-y+x & 1-y+x \\
3+y+x & 2+x & x \\
3+y+x & 2+x & x
\end{pmatrix}_{ij},
\label{eq:YDansatz}
\end{equation}
\begin{equation}
n_{i}^e=
diag\left(4+z+x,\ 2+x,\ x \right)
\label{eq:YEansatz}
\end{equation}
\begin{equation}
{\mathrm {where}}\ x=0,1,2,3, \ \ {\mathrm {and}} \ z=0,1.
\end{equation}
The integer parameter $x$ can be interpreted as the ambiguity of $\tan\beta\sim v_u/v_d$, {\it i.e.}, $m_b/m_t \sim \epsilon^x \cot\beta = \epsilon^x v_d/v_u$. On the one hand,the parameters $z$ determines the structure of the PMNS matrix. From those ansatzes, the CKM and PMNS mixing matrices can be obtained by $y$ and $z$ \cite{Harnik:2004yp},
\begin{equation}
V_{CKM} \sim
\begin{pmatrix}
1 & \epsilon^{1+y} & \epsilon^{3+y} \\
\epsilon^{1+y} & 1 & \epsilon^2 \\
\epsilon^{3+y} & \epsilon^2 & 1
\end{pmatrix}, \ \
V_{PMNS} \sim
\begin{pmatrix}
1 & \epsilon^z & \epsilon^z \\
\epsilon^z & 1 & 1 \\
\epsilon^z & 1 & 1 \\
\end{pmatrix}.
\label{eq:mixing mat}
\end{equation}
Those structures imply that the best prediction of mixing matrices is given when $y=0$ and $z=1$, and the other cases are semi-realistic one. The charge of flavor symmetry is strictly restricted.

\subsection{Higgs sector}
\label{sec:Higgs sector}
\ \ The Yukawa hierarchy can be obtained under the above ansatz, however, we must consider the Higgs sector. There are additional scalar degrees of freedom in MHDM and those predict additional physical states of Higgs fields. But such extra-Higgs fields are restricted by the experiments of neutral meson mixing \cite{Tanabashi:2018oca}. The mass difference depends on the mass of the mediator, therefore the extra-Higgs fields have to decouple at the high energy scale. Some previous works have discussed the decoupling of the extra-Higgs fields in MHDMs \cite{Branco:2011iw, Aranda:2000zf, Georgi:1978ri, Shanker:1981mj, McWilliams:1980kj, Escudero:2005hk}.

Before the discussion of the decoupling, we have to determine the discrete charge of the Higgs fields. If the extra-Higgs fields have the same discrete charge with the first generation, then not only the Yukawa couplings but also the mixing between the first generation and others in $\mu$-term are allowed. Those mixing terms induce the kinetic mixing and it is difficult to control the mass spectra of the Higgs fields. Thus, let us assume the extra-Higgs fields have a different discrete charge from the first generation. Note that the first generation of the Higgs fields corresponds to the MSSM-like Higgs fields based on this assumption since the Yukawa coupling with the extra-Higgs fields can be eliminated and the EWSB vacuum is given by $H_{u1}$ and $H_{d1}$. On the other hand, the different choice of the discrete charge may lead to the baryon number violation prohibited by the matter triality. The proton decay can occur in the case that the baryon and lepton numbers are broken simultaneously. In our model, the lepton number is violated by the interaction of the right-handed neutrino, therefore, the baryon number should be unbroken for the stability of the proton. The choice of the discrete charge can be determined uniquely from those requirements, the charge assignment within the extra-Higgs fields is given in Table \ref{tb:discrete charge within the extra-Higgs}, where the extra generation is denoted as Roman index. Under this discrete symmetry, the baryon number violating operator is prohibited up to mass dimension-5 (more detailed discussion, see Appendix \ref{sec:BLVop}).
\begin{table}[htb]
\centering
\caption{The representation of charged particle under matter triality within the extra-Higgs fields.}
\begin{tabular}{c|cccccccccccc}\hline\hline
& $Q$ & $\bar{U}$& $\bar{D}$& $L$& $\bar{E}$& $\bar{N}$ &$H_{u1}$ &$H_{d1}$ &$H_{ua}$ &$H_{da}$ \rule[-0pt]{0pt}{12pt} \\ \hline
$M_3$ & $1$ & $\omega^2$& $\omega$& $\omega$& $1$& $\omega$ &$\omega$ &$\omega^2$ &$\omega^2$ &$\omega$ \rule[-0pt]{0pt}{12pt} \\ \hline\hline
\end{tabular}
\label{tb:discrete charge within the extra-Higgs}
\end{table}

Then, let us discuss the mass spectra of the Higgs fields. The supersymmetric mass of the Higgs field comes from the $\mu$-term. Originally, the $\mu$-matrix with flavon is allowed in the superpotential, however, it is the only dimensionful parameter in the theory. Thus, its mass scale should correspond to the scale of underlying theory. Therefore, at the gravity scale, $\mu$-matrix has Planck scale, {\it i.e.}, $\mu_{\alpha\beta} \sim M_{Pl}$. Although the supersymmetric Higgs mass $\mu_{\alpha\beta}$ is responsible for the EW symmetry breaking, there is no reason why such mass relates to the EW scale, or why the scale is close to the SUSY breaking. This problem is called $\mu$-{\it problem}. In order to solve it, Giudice and Masiero proposed a mechanism which generates small mass scale related to the SUSY breaking scale $m_{soft}$ \cite{Giudice:1988yz}. Let $Z$ be a hidden sector chiral superfield, which is singlet under $G_{SM}\times \mathrm{U}(1)_X$. Then the non-minimal coupling in the K\"alher potential is introduced,
\begin{eqnarray}
g_{\alpha\beta}^{\mu}\int d^4\theta \frac{\bar{Z}}{M_{Pl}} \biggl\{\Theta[-X_{\alpha\beta}^\mu] \left(\frac{\bar{A}}{M_{Pl}} \right)^{-X_{\alpha\beta}^\mu} \nonumber +
\Theta[X_{\alpha\beta}^\mu]\left(\frac{A}{M_{Pl}} \right)^{X_{\alpha\beta}^\mu} \bigg\}H_{u\alpha}H_{d\beta} + h.c.,
\end{eqnarray}
where $X_{\alpha\beta}^\mu = X_{Hu\beta}+X_{Hd\beta}$. If we assume the gravity mediation SUSY breaking, then the F-term of the hidden sector superfields $\braket{F_Z}=m_{soft}M_{Pl}$, where $m_{soft}$ is soft SUSY breaking mass scale whose scale depends on the power of the flavon. After $Z$ is integrated out, the $\mu$-matrix which relates to the soft mass is given by,
\begin{equation}
\mu_{\alpha\beta} = g_{\alpha\beta}^\mu m_{soft} \epsilon^{|X_{Hu\alpha}+X_{Hd\beta}|} .
\end{equation}
Therefore, the effective $\mu$-matrix can be derived as
\begin{eqnarray}
\mu_{\alpha\beta} = \tilde{g}_{\alpha\beta}^\mu M_{Pl} \Theta\left[X_{Hu\alpha}+X_{Hd\beta}\right]\epsilon^{X_{Hu\alpha}+X_{Hd\beta}} +g_{\alpha\beta}^\mu m_{soft} \epsilon^{|X_{Hu\alpha}+X_{Hd\beta}|} .
\label{eq:muMAT}
\end{eqnarray}
Since the superpotential is holomorphic function of chiral superfields, its contribution depends on the sign of flavor charge while the contributions from the K\"ahler potential is always allowed\footnote{The prohibition by the holomorphy of the superpotential (SUSY zero) has been also discussed to generate the hierarchy of the scale. See Refs.\cite{Maekawa:2001uk, Maekawa:2001yh}.}. This means that the flavor symmetry is responsible for not only the hierarchy of fermion but also the hierarchy of Higgs fields. Note that the contribution from K\"alher potential to the operators with mass dimension greater than or equal to four are not significant because such coupling constants are suppressed by the Planck mass.

\section{Numerical analysis of flavor charge assignment}
\label{sec:res}
In this paper, we propose several examples of the concrete charge assignment which satisfies phenomenological and theoretical constraints as above mentioned. We analyze the charge assignments numerically, in order to obtain the realistic mass hierarchies. We assume that the ansatzes for Yukawa matrices with some parameters $x=0\sim 3$, $y=0,1$, and $z=0,1$, and analyze in the cases of the generation of Higgs fields $N_H = 1,2,3$. Furthermore, we parametrize the soft SUSY breaking scale, $M_{EW}\sim \epsilon^wm_{soft}$, where $w=1\sim 6$. We choose some X-charges as parameters under the above conditions Eq.(\ref{eq:YUansatz}), (\ref{eq:YDansatz}) and (\ref{eq:YEansatz}). Those parameters can be rewritten by using the following constraint in Table \ref{tb:DCtab1}.
\begin{table}[htb]
\caption{The constraints on the charge.}
\centering
\begin{tabular}{rll}\hline\hline
&$k_{Hu1}= -w - k_{Hd1}$ & \\
&$k_{Q2}=k_{Q1}-1-y$ & \\
&$k_{Q3}=k_{Q1}-3-y$ & \\
&$k_{U1}=-k_{Hu1}-k_{Q1}+8$ & \\
&$k_{U2}=k_{U1}-3+y$ & \\
&$k_{U3}=k_{U1}-5+y$ & \\
&$k_{D1}=-k_{Hd1}-k_{Q1}+4+x$ & \\
&$k_{D2}=k_{D1}-1+y$ & \\
&$k_{D3}=k_{D1}-1+y$ & \\
&$k_{E1}=-k_{Hd1}-k_{L1}+4+x+z$ & \\
&$k_{E2}=-k_{Hd1}-k_{L2}+2+x$ & \\
&$k_{E3}=-k_{Hd1}-k_{L3}+x$ & \\
\hline\hline
\end{tabular}
\label{tb:DCtab1}
\end{table}
For example, the charge of lepton doublet can be reduced as $X_{Li}=1+3k_{Li}$ $(k_{\Phi i}\in{\mathbb Z})$ due to the matter triality. On the other hand, in MHDM case, the extra-Higgs fields $H_{da}$ has to be different discrete charges to the first generation of Higgs field because of avoiding the mixing in mass matrix. Based on this assumption, the only one of the Higgs field have Yukawa coupling, the extra-Higgs fields decouple at the low energy. On the other hands, the flavor charge of the lepton doublet is also constrained in order to avoid the mixing between $L$ and the down-type extra-Higgs fields because those mixing induce the large lepton number violation. Thus, we require that the flavor charge of the lepton doublet should be negative so that $X_{Li}+X_{Hua}<0$. Then, the coupling constant is order $m_{soft}$ by the GM mechanism and the lepton number violation by the mixing between $L$ and $H_{da}$ can be suppressed (detail discussion is present in Appendix \ref{sec:BLVop}). Furthermore, we restrict the range of charge so that the perturbation is valid. The maximum and minimum are restricted as $|X_{max}|\leq 25$ and $|X_{max}/X_{min}|\leq 6$\footnote{The later constraint respects the hypercharge in SM, thus the result is conservative.}. The hypercharge normalization $k_Y$ is also arbitrary parameter, then we search the range $1\leq k_Y \leq 2$ for the solution. Again, note that the standard GUTs normalization is $k_Y=5/3$.

In such parameter space, the concrete examples of charge assignment are obtained (see Appendix \ref{sec:charge}). We obtain the two solutions and the 108 solutions with respect to $N_h=2,3$, while there is no solution for $N_h=1$. Let us show the concrete example of charge assignment. For the model No.77 in Table \ref{tb:DCA3-3}, the hypercharge normalization $k_Y=1$. The parameter $x$ relates to the ambiguity of $\tan\beta$, therefore this means
\begin{equation}
\tan\beta \sim \frac{m_t}{m_b}\epsilon^x \sim 10.
\end{equation}
The mass hierarchy and the mixing can be realized in this assignment. Under this symmetry, the ${\cal O}(1)$ factors of the Yukawa matrix are chosen as following:
\begin{equation}
Y^u =
\begin{pmatrix}
2.0 \epsilon ^8 & -2.0 \epsilon ^7 & 1.2 \epsilon ^5 \\
0.5 \epsilon ^5 & -2.1 \epsilon ^4 & 2.1 \epsilon ^2 \\
1.2 \epsilon ^3 & 0.5 \epsilon ^2 & 1.9 \\
\end{pmatrix}, \ \
Y^d =
\begin{pmatrix}
1.3 \epsilon ^7 & 1.0\epsilon ^6 & -1.0 \epsilon ^4 \\
-1.4 \epsilon ^6 & -1.7 \epsilon ^5 & 1.0 \epsilon ^3 \\
-0.8 \epsilon ^6 & 1.0\epsilon ^5 & -2.0 \epsilon ^3 \\
\end{pmatrix},
\end{equation}
\begin{equation}
Y^e =
\begin{pmatrix}
-2.1 \epsilon ^7 & -1. \epsilon ^6 & -0.5 \epsilon ^7 \\
2.1 \epsilon ^6 & 2.1 \epsilon ^5 & -0.5 \epsilon ^6 \\
-1.8 \epsilon ^3 & 0.5 \epsilon ^2 & -0.5 \epsilon ^3 \\
\end{pmatrix}.
\end{equation}
After diagonalization of mass matrices of quark and lepton, the mass spectra and CKM mixing can be derived as Table \ref{tb:FOcase1}.

\begin{table}[htb]
\centering
\caption{The fermion mass and mixing angle for the assignment No.77. The experimental values given in \cite{Tanabashi:2018oca}.}
\begin{tabular}{|c|c|c|} \hline
& experimental result & our result \rule[0mm]{0mm}{3mm} \\ \hline
\begin{tabular}{ccc}
$(m_u,$ & $m_c,$ & $m_t)/m_t$ \\
$(m_d,$ & $m_s,$ & $m_b)/m_b$ \\
$(m_e,$ & $m_\mu,$ & $m_\tau)/m_\tau$
\end{tabular}
&
\begin{tabular}{ccc}
$(1.3\times10^{-5},$ & $7.4\times10^{-3},$ & $1)$ \\
$(1.1\times10^{-3},$ & $2.3\times10^{-2},$ & $1)$ \\
$(2.9\times10^{-4},$ & $6.0\times10^{-2},$ & $1)$
\end{tabular}
&
\begin{tabular}{ccc}
$(5.6\times10^{-6},$ & $3.3\times10^{-3},$ & $1.0)$ \\
$(9.8\times10^{-4},$ & $2.4\times10^{-2},$ & $1.0)$ \\
$(4.3\times10^{-4},$ & $2.8\times10^{-2},$ & $1.0)$
\end{tabular} \\ \hline

$|V_{CKM}|$ & $\Bigg($
\begin{tabular}{ccc}
0.97 & 0.22 & 0.0039 \\
0.22 & 1.0 & 0.042 \\
0.0081 & 0.039 & 1.0
\end{tabular} $\Bigg)$
&$\Bigg($
\begin{tabular}{ccc}
0.97 & 0.25 & 0.0031 \\
0.25 & 0.97 & 0.049 \\
0.0095 & 0.049 & 1.0 \\
\end{tabular} $\Bigg)$\\ \hline
\end{tabular}
\label{tb:FOcase1}
\end{table}

The mass spectrum of the Higgs field is also determined by the flavor charge. The mixing between the first and the extra-Higgs fields in $\mu$-matrix are prohibited by $\mathrm{U}(1)_X$ symmetry since the extra-Higgs field have different discrete charges. Clearly, this matrix structure realizes the decoupling of extra-Higgs field, since the mixing between the first and the extra generation of the Higgs fields is prohibited by the discrete symmetry, {\it{i.e.}},
\begin{equation}
\mu_{ij}\sim\begin{pmatrix}
\epsilon^3m_{soft} & 0 & 0 \\
0 & \epsilon M_{Pl} & \epsilon^8M_{Pl} \\
0 & \epsilon M_{Pl} & \epsilon^8M_{Pl}\\
\end{pmatrix}_{ij}
\end{equation}
Then, the mass scale of the Higgs fields can be evaluated without ${\cal{O}}(1)$ factor,
\begin{equation}
m_H\sim \left(10^2,\ 10^{10},\ 10^{15}\right){\mathrm{GeV}}.
\label{eq:Higgsmassspectrum}
\end{equation}

The gauge anomaly coefficients are also calculated in this concrete charge assignment. The mixed anomalies with the SM gauge group are given by
\begin{equation}
({\cal A}_{CCX}, {\cal A}_{WWX}, {\cal A}_{YYX}, {\cal A}_{YXX}) = (108, 108, 108, 0).
\end{equation}
The GS field is necessary to cancel those anomalies.

Then, let us comment on our results. The parameters $y$ and $z$ are related to the quark and lepton mixing, as mentioned above Eq.(\ref{eq:mixing mat}). Although there are some solutions to realize the correct mixing patterns that obtain in $y=0$ and $z=1$, the neutrino sector also contributes to the PMNS matrix. The charge of right-handed neutrinos cannot be determined by the anomaly cancellation conditions, so we must consider those masses and flavor mixing. Secondly, we only obtain the solution $w=3$\footnote{Note that, we found that the solutions satisfy the constraints on the flavor charge when $w=3,6$ from the algebraic analysis. However, the solutions for $w=6$ are excluded by the condition of maximum and minimum charge assignments.}, this implies the soft SUSY breaking scale $m_{soft}\sim M_{EW}/\epsilon^3$, so the SUSY breaking scale is above a few ${\mathrm{TeV}}$. This result is consistent with the collider experiment. While those results are phenomenologically favored, the gauge coupling unification cannot be realized even if the correction of the extra-Higgs fields is included, because the normalization of the hypercharge $k_Y$ is around 1 while the usual GUT normalization is $5/3$. Naively, the small $k_Y$ comes from the tiny contribution of the flavor charge, therefore the normalization can be improved by the addition of the new field charged under $\mathrm{U}(1)_X$.

\section{Conclusion}
\label{sec:con}
In this paper, we analyze the multi-Higgs extension of SSM, where the $\mathrm{U}(1)_X$ flavor symmetry respecting matter triality is assigned. By gauging the flavor symmetry, the gauge anomaly cancellations are required. Furthermore, we require the hierarchical structure for the Yukawa matrix and also the supersymmetric mass matrix for the multi-Higgs fields via the Froggatt-Nielsen mechanism. Although various theoretical and phenomenological requirements strictly restrict the charge assignments, we confirm that the existence of the extra-Higgs fields plays a role to relax these constraints, where the gauge anomaly can be still canceled by the Green-Schwarz mechanism. In multi-Higgs doublet models, new scalar states couple to quarks and leptons as well as weak bosons, those fields mediate dangerous flavor changing neutral currents. To avoid this, we required such a charge assignment that the extra-Higgs fields decouple from the low energy. 

We numerically searched and found the concrete charge assignment which realizes the experimental values of fermion masses and the mixing. It is remarkable that the extra-Higgs fields have an intermediate-scale masses with the obtained charge assignments and the charge assignment requires TeV scale SUSY breaking. Note that the hierarchical structure of fermions and multi-Higgs fields are determined by the flavor symmetry. We emphasize that the large hierarchy in the Higgs spectrum is generated by two different sources, {\it {i.e.}}, the mass of the first generation of Higgs fields comes from the K\"ahler potential while the others come from the superpotential. In addition to the decoupling of the Higgs fields, the suppression of the mixing between the lepton doublet $L$ and down-type extra-Higgs fields $H_{da}$ by the same discrete charge can be also realized. Although the mixing by the bilinear term induces the lepton number violation, the contribution to the mixing only comes from the K\"ahler potential since the flavor charge of the lepton doublet can be negative and they have soft SUSY breaking scale $m_{soft}$. Because of the large $\mu$-term for the extra-Higgs fields, the mixing should be suppressed and controlled by the flavor symmetry. 

The mass spectrum of neutrino should be also determined by the flavor symmetry. The Yukawa coupling of the neutrino can be the main source to explain the tiny neutrino mass. In Refs.\cite{Dreiner:2006xw, Dreiner:2007vp}, the authors discussed the generation of neutrino mass in a similar context. However, there is an additional interaction between right-handed neutrinos in our model, which is allowed as the lepton number violating operator in the superpotential. Such specific interaction term has been discussed in the context of the $\mu\nu$SSM \cite{LopezFogliani:2005yw, Escudero:2008jg}. We will discuss the origin of neutrino mass in future works. In our model, since the R-parity is broken, the LSP is not stable and cannot be regarded as the candidate for the dark matter. The axion is another attractive candidate for the dark matter in the R-parity violating scenario \cite{Colucci:2018yaq}, where the model equips with the baryon triality for the stability of the proton instead of R-parity. We will also explore the supersymmetric axion model when matter triality is assigned.

\section*{Acknowledgments}
We would like to thank H. Abe for helpful discussions and advice.

\setcounter{section}{0}
\renewcommand\theHsection{\thesection}
\renewcommand{\thesection}{\Alph{section}}

\setcounter{equation}{0}
\renewcommand{\theequation}{\Alph{section}.\arabic{equation}}
\setcounter{figure}{0}
\renewcommand{\thefigure}{\Alph{section}.\arabic{figure}}
\setcounter{table}{0}
\renewcommand{\thetable}{\Alph{section}.\arabic{table}}

\section{Baryon/Lepton number violating operators}
\label{sec:BLVop}
In the supersymmetric SM, the baryon or lepton number violation terms are allowed, which induce the proton decay \cite{Goto:1998qg, Murayama:1994tc, Harnik:2004yp}. In SSM with right-handed neutrinos, those operators up to dimension-5 in the superpotential can be listed in Table \ref{tb:BL violating operators}.
\begin{table}[htb]
\begin{center}
\caption{The baryon/lepton number violation operator in the superpotential. Note that the dimension of operators refers to the mass dimension in terms of the Lagrangian after the integration of the Grassmann coordinates. The operator $W_Y$ means the Yukawa coupling in Eq.(\ref{eq:effectivesuperpotential}).}
\begin{tabular}{|c|c|c|c|c|c|c|}\hline
dim &	$\slashed{B}\ {\mathrm{and}}\ \slashed{L}$ & $\slashed{B} $	&	$\slashed{L}$\rule[-0pt]{0pt}{12pt}	\\ \hline
$2$	&	$$	&	$$	&	$\overline{N}$ \rule[-0pt]{0pt}{15pt}	\\	\hline
$3$	&	$$	&	$$	&	$\overline{NN},\ LH_u$	\rule[-0pt]{0pt}{15pt} \\	\hline
$4$	&	$$	&	$\overline{UDD}$	&	$\overline{NNN},\ LL\overline{E},\ LQ\overline{D},\ \overline{N}H_uH_d$	\rule[-0pt]{0pt}{15pt}	\\ \hline
\multirow{2}{*}{5} & \multirow{2}{*}{$QQQL,\ \overline{UUDE},\ \overline{UDDN}$} & \multirow{2}{*}{$QQQH_d$} & $Q\overline{UE}H_d,\ \overline{NNNN},\ LLH_uH_u$ \rule[-0pt]{0pt}{15pt} \\
& & & $LH_uH_uH_d,\ \overline{N}W_Y,\ \overline{NN}H_uH_d$ \rule[-0pt]{0pt}{15pt}\\ \hline
\end{tabular}
\label{tb:BL violating operators}
\end{center}
\end{table}
Now, let us assign the discrete symmetry in Table \ref{tb:discrete charge within the extra-Higgs}. Under the discrete symmetry, the Yukawa coupling for the first generation of the Higgs field and $\mu$-term are allowed, and the lepton number violating operators are given by
\begin{align}
& LH_{ua},\ \overline{NNN},\ \overline{N}H_{u1}H_{da}, Q\overline{UE}H_{da},\ LLH_{ua}H_{ub},\ \nonumber \\
& LH_{u1}H_{ua}H_{db},\ LH_{ua}H_{u1}H_{d1},\ \overline{NN}H_{ua}H_{d1},\ \overline{ND}QH_{da},\ \overline{NE}LH_{da}.
\end{align}
Due to the conservation of the baryon number\footnote{We can also check the baryon number conservation up to dimension-5 even if there exist the extra-Higgs fields and right-handed neutrino within the K\"ahler potential. The baryon or lepton violating operators in the K\"ahler potential is listed in Refs.\cite{Ibanez:1991pr, Dreiner:2005rd, Allanach:2003eb}.}, the proton is stable, however we need to carry out the basis rotation to canonicalize the kinetic term and eliminate the mixing between $L$ and $H_{da}$ \cite{Dreiner:2003hw} because those charge are same under $G_{SM}\times M_3$. After the canonicalization, the bilinear lepton number violation terms with the lepton doublet can be rotated away,
\begin{align}
H_{ua}
\begin{pmatrix}
\mu & \mu'
\end{pmatrix}_{aI}
\begin{pmatrix}
H_d \\
L
\end{pmatrix}_I
&=H_{ua}
\begin{pmatrix}
\tilde{\mu} & 0
\end{pmatrix}_{aI}
\begin{pmatrix}
\tilde{H_d} \\
\tilde{L}
\end{pmatrix}_I,\\
{\mathrm{where}}\
\begin{pmatrix}
\tilde{H_d} \\
\tilde{L}
\end{pmatrix}_I
&=U^{(K)}_{IJ}
\begin{pmatrix}
{H_d} \\
{L}
\end{pmatrix}_J.
\end{align}
The capital index run over the generation of the lepton doublet and the extra-Higgs fields. For simplicity, let us assume the extra-Higgs field is one-generation, then the bilinear coupling constant $K_{aI}$ can be reduced to $K_I$,
\begin{equation}
U_{IJ}^{(K)\dagger} = \frac{|\mu|}{{\cal{M}}}
\begin{pmatrix}
1 & \left(\dfrac{\mu'_i}{\mu}\right) \\
-\left(\dfrac{\mu'_j}{\mu}\right)^* & \dfrac{{\mu'}_j\mu'_i}{\mu'^2}\left(1-\dfrac{{\cal{M}}}{\mu} \right)+\dfrac{{\cal{M}}}{|\mu|}\delta_{ij}
\end{pmatrix}_{IJ} ,
\end{equation}
where ${\cal{M}}=\sqrt{{\mu'}_i^*\mu'_i+\mu^*\mu}$ and ${\mu'}^2=\sqrt{{\mu'}_i^*\mu'_i}$.

In our scheme, the bilinear lepton number violation term is induced from the K\"ahler potential by Giudice-Masiero mechanism \cite{Giudice:1988yz} while the term coupling with the flavon is originally allowed under the symmetry. Now we assume that the extra-Higgs fields can be heavy and those masses are around intermediate scale, therefore if the mixing $\mu'_i$ is enough small, then the lepton number violation terms induced by the mixing should be small, {\it{i.e.}}, $\mu' \ll \mu $, then
\begin{align}
H_d &=\dfrac{|\mu|}{{\cal{M}}}\left( \tilde{H_d} + \dfrac{\mu'_i}{\mu}\tilde{L_i} \right)\ \sim \ \tilde{H_d}.\\
L_i &=\dfrac{|\mu|}{{\cal{M}}}\left( \dfrac{\mu'_i}{\mu}\tilde{H_d} -\left( \dfrac{{\mu'}_j\mu'_i}{\mu'^2}\left(1-\dfrac{{\cal{M}}}{\mu} \right)+\dfrac{{\cal{M}}}{|\mu|}\delta_{ij}\right)\tilde{L_j} \right)\ \sim \ \tilde{L_i}.
\end{align}


\section{The charge assignments}
\label{sec:charge}
The parameters which provide the concrete charge assignments are listed in Table \ref{tb:DCA2} for $N_h=2$ and Table \ref{tb:DCA3-1}, \ref{tb:DCA3-2}, and \ref{tb:DCA3-3} for $N_h=3$. Those parameters can be rewritten into the X-charge by using the condition of the Yukawa matrices. We obtain the two solutions for $N_h=2$ and 108 solutions for $N_h=3$.

\begin{footnotesize}
\begin{table}[htb]
\caption{The parameters of the charge assignment ($N_h=2$).}
\centering
\begin{tabular}{ccccccccccccc}\hline\hline
No.& $k_{Q1}$ & $k_{L1}$ & $k_{L2}$ & $k_{L3}$ & $k_{Hu2}$ & $k_{Hd1}$ & $k_{Hd2}$ & $x$ & $y$ & $z$ & $w$ & $k_Y$ \\ \hline \hline
1& 6 & -6 & -7 & -6 & 5 & 3 & 4 & 1 & 0 & 0 & 3 & 31/30 \\
2& 7 & -5 & -8 & -5 & 4 & 4 & 4 & 2 & 1 & 1 & 3 & 67/66 \\ \hline\hline
\end{tabular}
\label{tb:DCA2}
\end{table}
\end{footnotesize}

\begin{footnotesize}
\begin{table}[htb]
\caption{The parameters of the charge assignment ($N_h=3$).}
\centering
\begin{tabular}{ccccccccccccccc}\hline\hline
No.& $k_{Q1}$ & $k_{L1}$ & $k_{L2}$ & $k_{L3}$ & $k_{Hu2}$ & $k_{Hu3}$ & $k_{Hd1}$ & $k_{Hd2}$ & $k_{Hd3}$ & $x$ & $y$ & $z$ & $w$ & $k_Y$ \\ \hline \hline
1 & 6 & -5 & -8 & -8 & -5 & 2 & 2 & 3 & 8 & 0 & 0 & 1 & 3 & 19/18 \\
2 & 6 & -6 & -8 & -8 & -4 & 3 & 2 & 3 & 7 & 0 & 0 & 0 & 3 & 19/18 \\
3 & 6 & -6 & -8 & -8 & -2 & 2 & 2 & 4 & 5 & 0 & 0 & 0 & 3 & 19/18 \\
4 & 6 & -5 & -8 & -6 & 3 & 4 & 2 & 1 & 4 & 0 & 1 & 1 & 3 & 65/54 \\
5 & 6 & -6 & -8 & -6 & 4 & 4 & 2 & 2 & 3 & 0 & 1 & 0 & 3 & 65/54 \\
6 & 6 & -4 & -8 & -5 & 3 & 3 & 2 & 2 & 2 & 0 & 1 & 1 & 3 & 61/54 \\
7 & 6 & -5 & -8 & -4 & 2 & 3 & 2 & 2 & 3 & 0 & 1 & 1 & 3 & 61/54 \\
8 & 6 & -5 & -7 & -7 & -6 & 2 & 2 & 5 & 8 & 1 & 0 & 0 & 3 & 31/30 \\
9 & 6 & -5 & -7 & -7 & 3 & 4 & 2 & -5 & 7 & 1 & 0 & 0 & 3 & 31/30 \\
10 & 6 & -5 & -7 & -7 & 3 & 4 & 2 & 2 & 3 & 0 & 1 & 1 & 3 & 65/54 \\
11 & 6 & -5 & -7 & -7 & 4 & 4 & 2 & 3 & 4 & 1 & 1 & 0 & 3 & 37/30 \\
12 & 6 & -6 & -7 & -5 & 3 & 4 & 2 & 1 & 3 & 0 & 1 & 0 & 3 & 61/54 \\
13 & 6 & -4 & -7 & -3 & 2 & 2 & 2 & 3 & 3 & 1 & 1 & 1 & 3 & 11/10 \\
14 & 6 & -3 & -6 & -6 & -5 & -1 & 2 & 7 & 7 & 2 & 0 & 1 & 3 & 67/66 \\
15 & 6 & -3 & -6 & -6 & -3 & -2 & 2 & 6 & 7 & 2 & 0 & 1 & 3 & 67/66 \\
16 & 6 & -3 & -6 & -6 & -2 & -2 & 2 & 4 & 8 & 2 & 0 & 1 & 3 & 67/66 \\
17 & 6 & -3 & -6 & -6 & 2 & 2 & 2 & -4 & 8 & 2 & 0 & 1 & 3 & 67/66 \\
18 & 6 & -4 & -6 & -6 & 2 & 2 & 2 & -3 & 8 & 2 & 0 & 0 & 3 & 67/66 \\
19 & 6 & -6 & -6 & -6 & 2 & 5 & 2 & 1 & 3 & 0 & 1 & 0 & 3 & 61/54 \\
20 & 6 & -5 & -6 & -6 & 3 & 3 & 2 & 1 & 3 & 0 & 1 & 1 & 3 & 61/54 \\
21 & 6 & -4 & -6 & -6 & 3 & 3 & 2 & 2 & 4 & 1 & 1 & 1 & 3 & 7/6 \\
22 & 6 & -5 & -6 & -6 & 3 & 4 & 2 & 3 & 3 & 1 & 1 & 0 & 3 & 7/6 \\
23 & 6 & -5 & -6 & -5 & 4 & 4 & 2 & -2 & 3 & 0 & 1 & 0 & 3 & 19/18 \\
24 & 6 & -5 & -6 & -4 & 2 & 2 & 2 & 2 & 2 & 0 & 1 & 1 & 3 & 19/18 \\
25 & 6 & -3 & -6 & -3 & 2 & 2 & 2 & 2 & 2 & 1 & 1 & 1 & 3 & 31/30 \\
26 & 6 & -5 & -5 & -5 & -7 & 4 & 2 & 5 & 6 & 0 & 1 & 1 & 3 & 19/18 \\
27 & 6 & -5 & -5 & -5 & -1 & 4 & 2 & 2 & 3 & 0 & 1 & 1 & 3 & 19/18 \\
28 & 6 & -6 & -5 & -5 & 2 & 4 & 2 & 1 & 2 & 0 & 1 & 0 & 3 & 19/18 \\
29 & 6 & -5 & -5 & -5 & 2 & 4 & 2 & 2 & 3 & 1 & 1 & 0 & 3 & 11/10 \\
30 & 6 & -4 & -4 & -4 & 2 & 2 & 2 & 1 & 3 & 1 & 1 & 1 & 3 & 31/30 \\
31 & 6 & -3 & -4 & -4 & 2 & 2 & 2 & 2 & 4 & 2 & 1 & 1 & 3 & 71/66 \\
32 & 6 & -4 & -4 & -4 & 2 & 3 & 2 & 3 & 3 & 2 & 1 & 0 & 3 & 71/66 \\
33 & 6 & -4 & -4 & -4 & 3 & 3 & 2 & -2 & 4 & 1 & 1 & 1 & 3 & 31/30 \\
34 & 7 & -6 & -8 & -8 & -6 & 3 & 3 & 4 & 8 & 1 & 1 & 0 & 3 & 31/30 \\
35 & 7 & -6 & -8 & -8 & -3 & 3 & 3 & 2 & 7 & 1 & 1 & 0 & 3 & 31/30 \\
36 & 7 & -6 & -8 & -8 & -1 & 2 & 3 & 2 & 6 & 1 & 1 & 0 & 3 & 31/30
\\\hline\hline
\end{tabular}
\label{tb:DCA3-1}
\end{table}
\end{footnotesize}

\begin{footnotesize}
\begin{table}[htb]
\caption{The parameters of the charge assignment ($N_h=3$) {\it{conti.}}.}
\centering
\begin{tabular}{ccccccccccccccc}\hline\hline
No.& $k_{Q1}$ & $k_{L1}$ & $k_{L2}$ & $k_{L3}$ & $k_{Hu2}$ & $k_{Hu3}$ & $k_{Hd1}$ & $k_{Hd2}$ & $k_{Hd3}$ & $x$ & $y$ & $z$ & $w$ & $k_Y$ \\ \hline \hline
37 & 6 & -6 & -8 & -8 & 2 & 5 & 3 & 1 & 1 & 0 & 0 & 0 & 3 & 19/18 \\
38 & 6 & -5 & -8 & -8 & 3 & 3 & 3 & 1 & 1 & 0 & 0 & 1 & 3 & 19/18 \\
39 & 7 & -6 & -8 & -8 & 5 & 5 & 3 & -7 & 6 & 1 & 1 & 0 & 3 & 31/30 \\
40 & 6 & -6 & -8 & -7 & -8 & 4 & 3 & 5 & 7 & 0 & 0 & 1 & 3 & 19/18 \\
41 & 6 & -6 & -8 & -7 & -7 & 5 & 3 & 2 & 8 & 0 & 0 & 1 & 3 & 19/18 \\
42 & 6 & -7 & -8 & -7 & -6 & 6 & 3 & 2 & 7 & 0 & 0 & 0 & 3 & 19/18 \\
43 & 6 & -7 & -8 & -7 & -3 & 6 & 3 & 1 & 5 & 0 & 0 & 0 & 3 & 19/18 \\
44 & 6 & -7 & -8 & -7 & 4 & 4 & 3 & -2 & 3 & 0 & 0 & 0 & 3 & 19/18 \\
45 & 6 & -5 & -8 & -6 & 3 & 3 & 3 & 1 & 2 & 1 & 0 & 0 & 3 & 31/30 \\
46 & 6 & -5 & -8 & -5 & -6 & 4 & 3 & 3 & 7 & 1 & 0 & 1 & 3 & 31/30 \\
47 & 6 & -5 & -8 & -5 & -2 & 4 & 3 & 1 & 5 & 1 & 0 & 1 & 3 & 31/30 \\
48 & 6 & -6 & -8 & -5 & 2 & 3 & 3 & 2 & 2 & 1 & 0 & 0 & 3 & 31/30 \\
49 & 6 & -6 & -8 & -5 & 3 & 4 & 3 & -2 & 4 & 1 & 0 & 0 & 3 & 31/30 \\
50 & 7 & -5 & -7 & -7 & -6 & 2 & 3 & 5 & 8 & 2 & 1 & 0 & 3 & 67/66 \\
51 & 6 & -4 & -7 & -7 & 2 & 3 & 3 & 1 & 2 & 1 & 0 & 1 & 3 & 31/30 \\
52 & 7 & -5 & -7 & -7 & 3 & 4 & 3 & -5 & 7 & 2 & 1 & 0 & 3 & 67/66 \\
53 & 6 & -6 & -7 & -6 & -8 & 5 & 3 & 4 & 8 & 1 & 0 & 0 & 3 & 31/30 \\
54 & 6 & -6 & -7 & -6 & -7 & 5 & 3 & 4 & 7 & 1 & 0 & 0 & 3 & 31/30 \\
55 & 6 & -6 & -7 & -6 & -6 & 5 & 3 & 4 & 6 & 1 & 0 & 0 & 3 & 31/30 \\
56 & 6 & -6 & -7 & -6 & -5 & 5 & 3 & 4 & 5 & 1 & 0 & 0 & 3 & 31/30 \\
57 & 6 & -5 & -7 & -6 & -4 & 4 & 3 & 3 & 5 & 1 & 0 & 1 & 3 & 31/30 \\
58 & 6 & -6 & -7 & -6 & -4 & 5 & 3 & 4 & 4 & 1 & 0 & 0 & 3 & 31/30 \\
59 & 6 & -6 & -7 & -6 & -3 & 5 & 3 & 3 & 4 & 1 & 0 & 0 & 3 & 31/30 \\
60 & 6 & -6 & -7 & -6 & -2 & 5 & 3 & 2 & 4 & 1 & 0 & 0 & 3 & 31/30 \\
61 & 6 & -6 & -7 & -6 & -1 & 5 & 3 & 1 & 4 & 1 & 0 & 0 & 3 & 31/30 \\
62 & 6 & -5 & -7 & -6 & 2 & 2 & 3 & 1 & 3 & 1 & 0 & 1 & 3 & 31/30 \\
63 & 6 & -4 & -7 & -6 & 2 & 2 & 3 & 3 & 3 & 2 & 0 & 1 & 3 & 71/66 \\
64 & 6 & -6 & -7 & -6 & 2 & 5 & 3 & -2 & 4 & 1 & 0 & 0 & 3 & 31/30 \\
65 & 6 & -5 & -7 & -6 & 3 & 3 & 3 & -2 & 4 & 1 & 0 & 1 & 3 & 31/30 \\
66 & 6 & -6 & -7 & -6 & 3 & 5 & 3 & -3 & 4 & 1 & 0 & 0 & 3 & 31/30 \\
67 & 6 & -6 & -7 & -6 & 4 & 5 & 3 & -4 & 4 & 1 & 0 & 0 & 3 & 31/30 \\
68 & 6 & -6 & -7 & -6 & 5 & 5 & 3 & -5 & 4 & 1 & 0 & 0 & 3 & 31/30 \\
69 & 6 & -4 & -7 & -5 & 2 & 3 & 3 & 1 & 3 & 2 & 0 & 0 & 3 & 67/66 \\
70 & 6 & -4 & -7 & -4 & -7 & 3 & 3 & 6 & 6 & 2 & 0 & 1 & 3 & 67/66 \\
71 & 6 & -4 & -7 & -4 & -3 & 3 & 3 & 4 & 4 & 2 & 0 & 1 & 3 & 67/66 \\
72 & 6 & -5 & -7 & -4 & 2 & 2 & 3 & 2 & 3 & 2 & 0 & 0 & 3 & 67/66
\\\hline\hline
\end{tabular}
\label{tb:DCA3-2}
\end{table}
\end{footnotesize}
\begin{footnotesize}

\begin{table}[htb]
\caption{The parameters of the charge assignment ($N_h=3$) {\it{conti.2}}.}
\centering
\begin{tabular}{ccccccccccccccc}\hline\hline
No.& $k_{Q1}$ & $k_{L1}$ & $k_{L2}$ & $k_{L3}$ & $k_{Hu2}$ & $k_{Hu3}$ & $k_{Hd1}$ & $k_{Hd2}$ & $k_{Hd3}$ & $x$ & $y$ & $z$ & $w$ & $k_Y$ \\ \hline \hline
73 & 7 & -4 & -6 & -6 & -1 & -1 & 3 & 4 & 7 & 3 & 1 & 0 & 3 & 1 \\
74 & 6 & -3 & -6 & -6 & 2 & 2 & 3 & 2 & 2 & 2 & 0 & 1 & 3 & 67/66 \\
75 & 6 & -4 & -6 & -5 & -1 & 3 & 3 & 2 & 4 & 2 & 0 & 1 & 3 & 67/66 \\
76 & 6 & -5 & -6 & -5 & -1 & 4 & 3 & 3 & 3 & 2 & 0 & 0 & 3 & 67/66 \\
77 & 6 & -4 & -5 & -4 & 3 & 3 & 3 & -2 & 5 & 3 & 0 & 0 & 3 & 1 \\
78 & 7 & -5 & -8 & -8 & 4 & 4 & 4 & 1 & 2 & 2 & 1 & 0 & 3 & 71/66 \\
79 & 7 & -6 & -8 & -7 & -5 & 5 & 4 & 2 & 6 & 1 & 1 & 1 & 3 & 31/30 \\
80 & 7 & -7 & -8 & -7 & -4 & 6 & 4 & 2 & 5 & 1 & 1 & 0 & 3 & 31/30 \\
81 & 7 & -7 & -8 & -7 & -2 & 6 & 4 & 1 & 4 & 1 & 1 & 0 & 3 & 31/30 \\
82 & 7 & -6 & -8 & -7 & 2 & 5 & 4 & 1 & 3 & 2 & 1 & 0 & 3 & 71/66 \\
83 & 7 & -5 & -8 & -7 & 3 & 3 & 4 & 1 & 3 & 2 & 1 & 1 & 3 & 71/66 \\
84 & 7 & -7 & -8 & -7 & 3 & 4 & 4 & 1 & 1 & 1 & 1 & 0 & 3 & 31/30 \\
85 & 7 & -7 & -8 & -7 & 4 & 5 & 4 & -3 & 3 & 1 & 1 & 0 & 3 & 31/30 \\
86 & 7 & -5 & -8 & -6 & 4 & 4 & 4 & -2 & 3 & 2 & 1 & 0 & 3 & 67/66 \\
87 & 7 & -5 & -8 & -5 & -8 & 4 & 4 & 4 & 8 & 2 & 1 & 1 & 3 & 67/66 \\
88 & 7 & -5 & -8 & -5 & -7 & 4 & 4 & 4 & 7 & 2 & 1 & 1 & 3 & 67/66 \\
89 & 7 & -5 & -8 & -5 & -6 & 4 & 4 & 4 & 6 & 2 & 1 & 1 & 3 & 67/66 \\
90 & 7 & -5 & -8 & -5 & -5 & 4 & 4 & 4 & 5 & 2 & 1 & 1 & 3 & 67/66 \\
91 & 7 & -5 & -8 & -5 & -4 & 4 & 4 & 4 & 4 & 2 & 1 & 1 & 3 & 67/66 \\
92 & 7 & -5 & -8 & -5 & -3 & 4 & 4 & 3 & 4 & 2 & 1 & 1 & 3 & 67/66 \\
93 & 7 & -5 & -8 & -5 & -2 & 4 & 4 & 2 & 4 & 2 & 1 & 1 & 3 & 67/66 \\
94 & 7 & -5 & -8 & -5 & -1 & 4 & 4 & 1 & 4 & 2 & 1 & 1 & 3 & 67/66 \\
95 & 7 & -5 & -8 & -5 & 2 & 2 & 4 & 2 & 2 & 2 & 1 & 1 & 3 & 67/66 \\
96 & 7 & -5 & -8 & -5 & 2 & 4 & 4 & -2 & 4 & 2 & 1 & 1 & 3 & 67/66 \\
97 & 7 & -5 & -8 & -5 & 3 & 4 & 4 & -3 & 4 & 2 & 1 & 1 & 3 & 67/66 \\
98 & 7 & -5 & -8 & -5 & 4 & 4 & 4 & -4 & 4 & 2 & 1 & 1 & 3 & 67/66 \\
99 & 7 & -5 & -7 & -7 & 3 & 4 & 4 & 1 & 1 & 2 & 1 & 0 & 3 & 67/66 \\
100 & 7 & -6 & -7 & -6 & -8 & 5 & 4 & 5 & 7 & 2 & 1 & 0 & 3 & 67/66 \\
101 & 7 & -5 & -7 & -6 & -7 & 4 & 4 & 5 & 6 & 2 & 1 & 1 & 3 & 67/66 \\
102 & 7 & -5 & -7 & -6 & -1 & 4 & 4 & 2 & 3 & 2 & 1 & 1 & 3 & 67/66 \\
103 & 7 & -6 & -7 & -6 & 2 & 4 & 4 & 1 & 2 & 2 & 1 & 0 & 3 & 67/66 \\
104 & 7 & -5 & -7 & -6 & 2 & 4 & 4 & 2 & 3 & 3 & 1 & 0 & 3 & 19/18 \\
105 & 7 & -5 & -7 & -4 & 2 & 3 & 4 & 1 & 3 & 3 & 1 & 0 & 3 & 1 \\
106 & 7 & -4 & -6 & -5 & 2 & 2 & 4 & 1 & 3 & 3 & 1 & 1 & 3 & 1 \\
107 & 7 & -4 & -6 & -5 & 3 & 3 & 4 & -2 & 4 & 3 & 1 & 1 & 3 & 1 \\
108 & 7 & -5 & -6 & -5 & 4 & 4 & 4 & -3 & 4 & 3 & 1 & 0 & 3 & 1 \\\hline\hline
\end{tabular}
\label{tb:DCA3-3}
\end{table}
\end{footnotesize}

\clearpage

\bibliography{reference}

\bibliographystyle{JHEP}

\end{document}